\documentstyle[12pt,epsfig]{article}
\title{\bf Are High Energy Heavy Ion Collisions similar to a Little Bang, 
or just a very nice Firework? }
 \author{ E.V.~Shuryak\\State University of New York, 
       Stony Brook, NY 11794, USA}
\begin{document} 
\maketitle
\begin{abstract}
The talk is a brief overview of what we recently learned about excited 
hadronic matter from heavy ion collisions. 
The central issue is that  the systems produced do exhibit macroscopic
behavior, it flows and we
start getting some idea about
 its effective Equation of State (EoS). 
More specifically, we concentrate on elliptic flow, from SPS to RHIC
energies, as well as on
 particle composition and fluctuations. Note
that  a pressure and
 the rate of fluctuation relaxation
(discussed at the end) are ultimately a measure
of a collision rate in the system we would like to understand.

\end{abstract}

\section{Flows}
\subsection{QCD phase transition and flows}
We start with general questions, such as: Do we produce excited
matter with  sufficiently large scattering rate able to ensure local
equilibration? If so, what is its EoS and whether it is close to 
results obtained in lattice simulations? 
How one can tell the Bang from a Fizzle, experimentally?
There are 3 effects one can discuss: (i) longitudinal
work; (ii) radial transverse flow; and (iii) elliptic 
flow. We address two last ones below.

Before we go to specific description of concepts and data on flow,
let us discuss in general why we think that rather different
phenomena at AGS/SPS and RHIC energy domain can nevertheless be
described in a unified way
by hydro+cascade model. Those types of models are the only ones
known, which can incorporate correctly the fact that hadrons and
partons
 live in different vacua, separated by significant ``bag term'' in
EoS.
This phenomenon generates soft ``mixed phase'', in which energy
density $\epsilon$ 
grows but temperature T and pressure p do not. Purely cascade models
with partons/hadrons miss this central point, and therefore
have very unrealistic EoS. 

  The Hydro-to-Hadrons  Model \cite{TLS} include hydro plus
 transfer to hadronic cascade
(RQMD),
  in order not to worry about  freeze-out of
    different species, resonance decays etc.
   The transfer is smooth enough because the effective
 EoS of RQMD and our
  hadronic matter is about the same.

 At SPS the evolution starts close to
rather soft ``mixed phase'' (as lattice thermodynamics tells us),
then proceeding to
stiffer pion gas: therefore most of the radial expansion is pion-driven
and happens very late. There are many 
 proves of that: one \cite{Sorge_omega} is that
 $\Omega^-$ which participate little in
hadronic rescattering
 practically do not have it.

 Let us first characterize flows in general.
  At RHIC we start well above the QCD
phase transition, and so expect the so called ``QGP push'',  then  softer
mixed phase, and finally a stiff hadronic phase again.  (Now $\Omega^-$ 
is expected to flow  more!)

 In non-central collisions at SPS
 the initial almond for $b\sim R$ collisions
retains basically the same almond shape: matter does not move till
the final velocity $v_t\approx 0.5$ is only obtained close to the end
of the expansion.
 At RHIC  the initial almond is transformed into a completely different
object called the {\em ``nutcracker''} \cite{TS_nut},
which consists of two separated shells of matter and a small ``nut''
in the center. It happens 
 by the time 8-10 fm/c, and then shells continue to move
out in hadronic phase, slowly dissolving.

  Radial flow is usually characterized by the slope parameter T:
$dN/dp_t^2dy\sim exp(-m_t/T), m_t^2=p_t^2+m^2$.
 T is $not$ temperature: it incorporates random thermal motion
and collective transverse velocity. Its prediction for various EoS
(marked by latent heat, say LH8 means latent heat 800 $MeV/fm^3$,
see fig.\ref{fig_1}a)
is shown in the following figure , for protons and pions versus
basically collision energy expressed in terms of multiplicity,
see  fig.\ref{fig_1}b.
One can see that different EoS show different growth, although picture 
is rather simple: the softer the EoS the less flow.
\begin{figure}[t]
\begin{center}
   \includegraphics[width=2.5in]{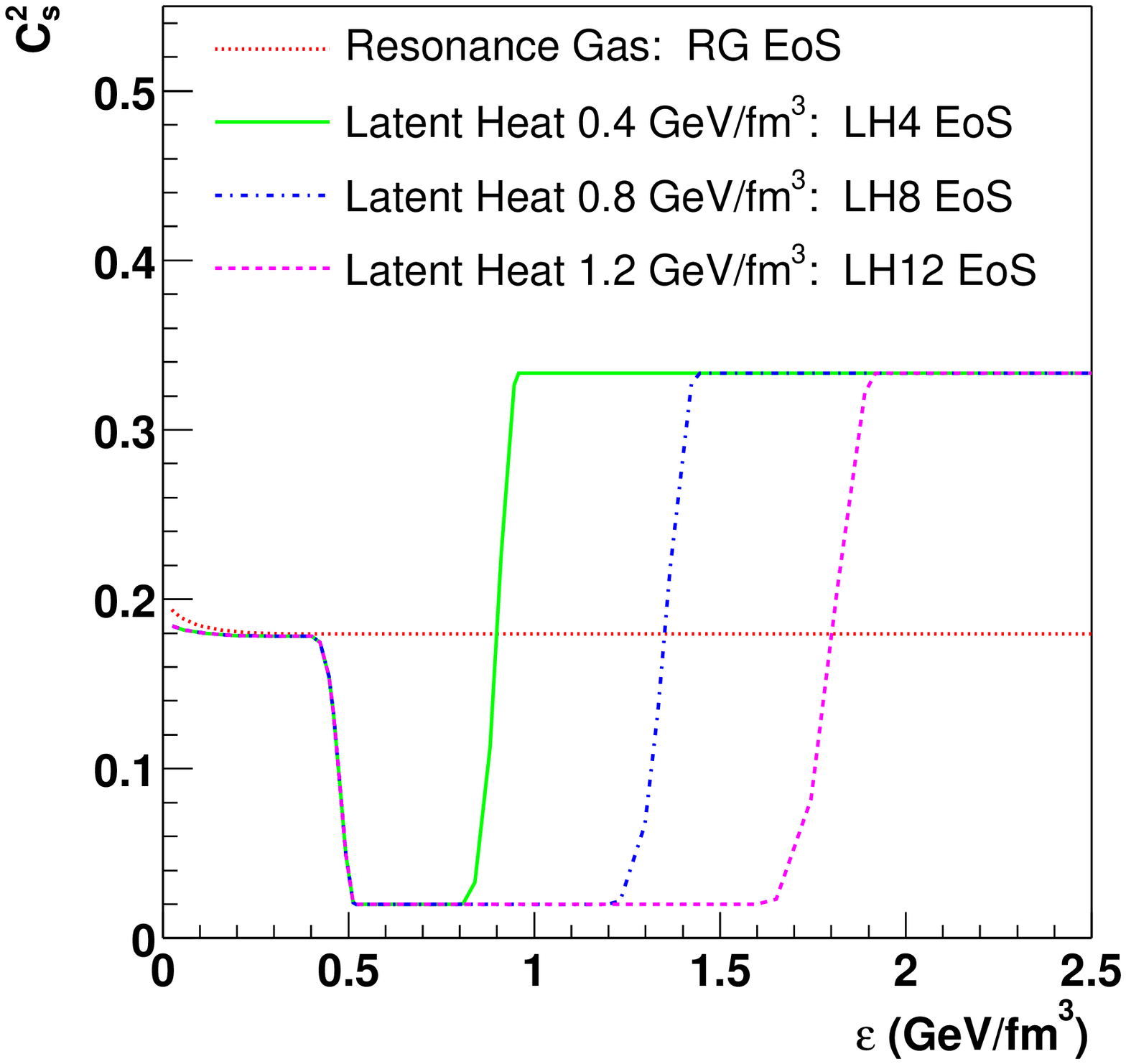}
   \includegraphics[width=2.5in]{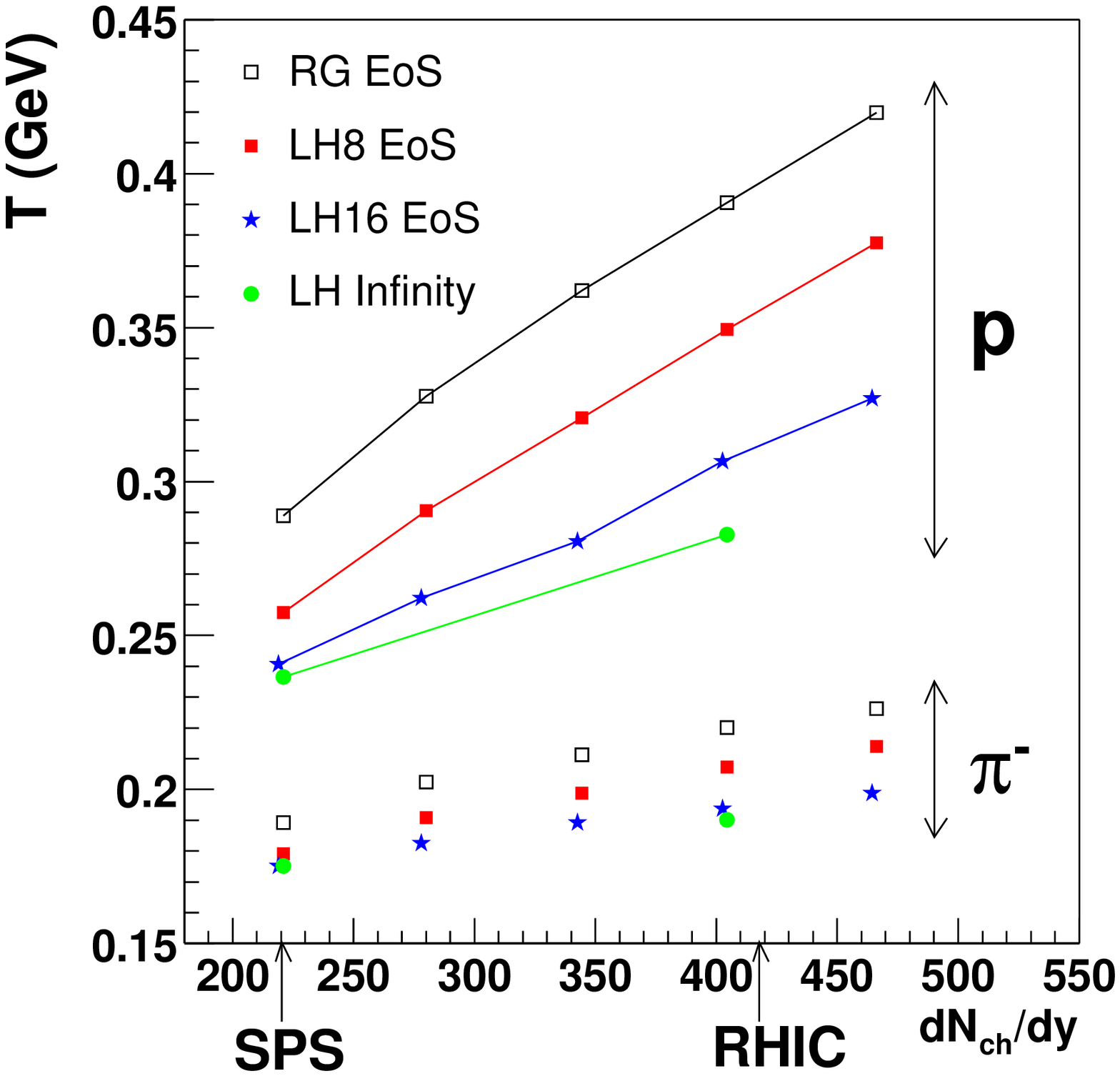}
\end{center}
   \caption[]{
      \label{fig_1}
The speed of sound for a set of EoS used (a) 
and resulting transverse mass slope  T (b) versus mid-rapidity (y=0)  charged particle
multiplicity,
for AuAu collisions with b=6.
   }
\end{figure}
 
\subsection{Elliptic flow}

 If there is no correlation between space and
momentum, there is no elliptic flow. For example, $any$ model like \cite{Leonidov_etal}
in which transverse momentum increase in AA is due to
initial state interaction, like in pA collision,
cannot have elliptic flow. Indeed HIJING parton model \cite{HIJING}
 without rescatterng (an example of the model which produce a
``firework''
type final state)
 has nearly zero (actually slightly negative) $v_2$. String models
like UrQMD  \cite{UrQMD_Bleicher} and RQMD itself
also do not produce pressure
at early time, and predicted respectively $smaller$ $v_2$ at RHIC than at SPS
 Those are eliminated, as soon as the first STAR data \cite{STAR} have shown that
at RHIC  $v_2$ is in fact twice larger than at
SPS!

But  hydro knows about geometry of the excited system:
the pressure gradient is mostly along the narrow part of the almond.
 The  elliptic flow is quantified experimentally
by  the elliptic flow
parameter $V_{2}=\langle cos(2\phi) \rangle$

The {\em energy dependence} of $V_{2}$ does not appear
to be simple (in contrast to radial flow). Furthermore,
 we  have found 
that if one is making EoS softer the $v_2$  decreases non-monotonously,
first decreasing and then increasing again. It means for a given
experimental value of  $V_{2}$
there are two possible  scenarios, we called 
the ``QGP push'' and  ``burning
almond''. In the first case the initial almond transfers to
nutcracker, in which spatial anisotropy decreases and even change sign.
In the second, the almond dries out, and  spatial anisotropy actually
grows. So, in order to answer the question whether
 the ``QGP push'' scenario we are waiting for is or is not the case,
some further studies are needed.
In particular, the scan in RHIC energies downward would be very
useful.

\begin{figure}[t]
\begin{center}
   \includegraphics[width=2.5in]{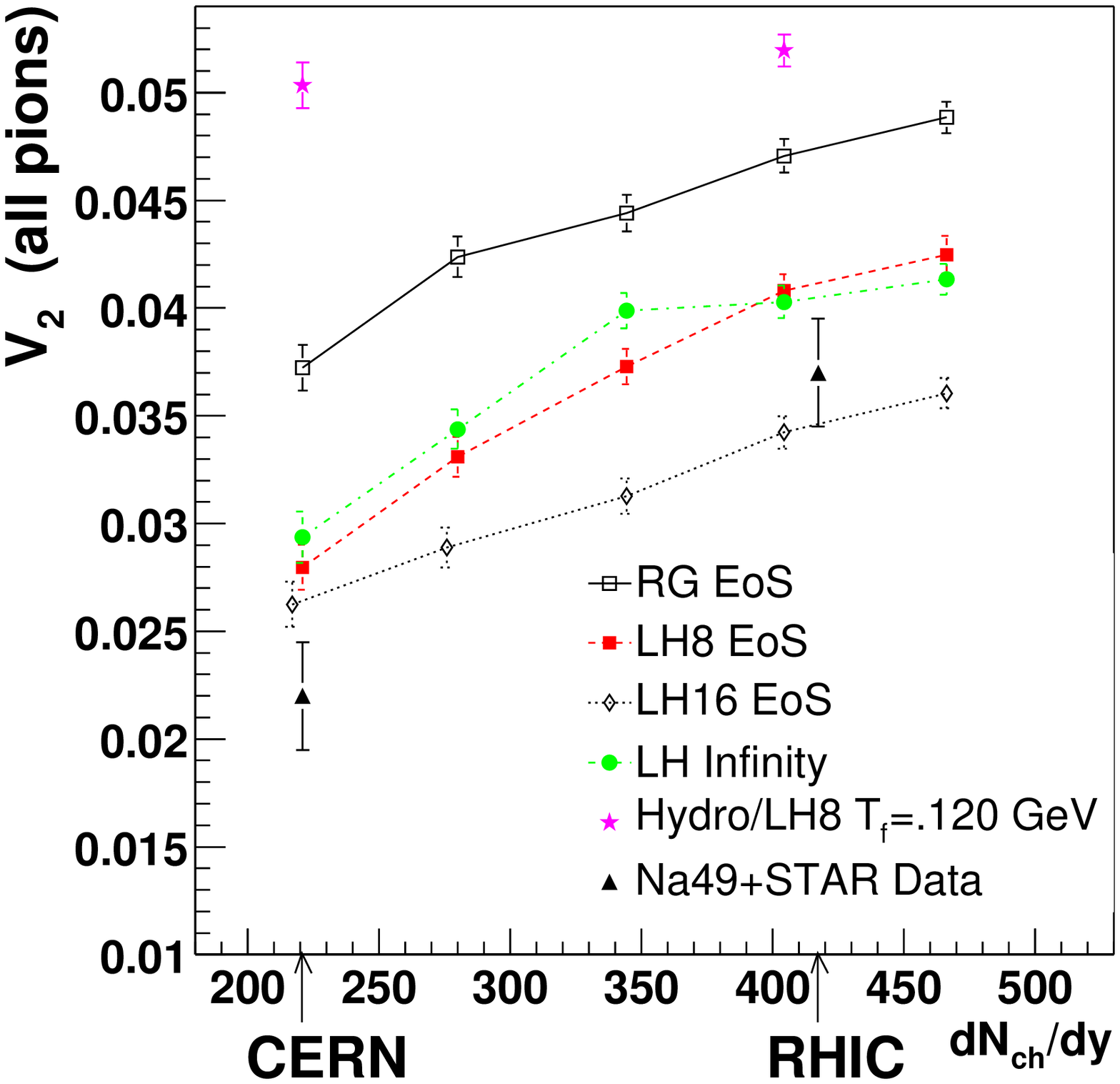}
   \includegraphics[width=2.5in]{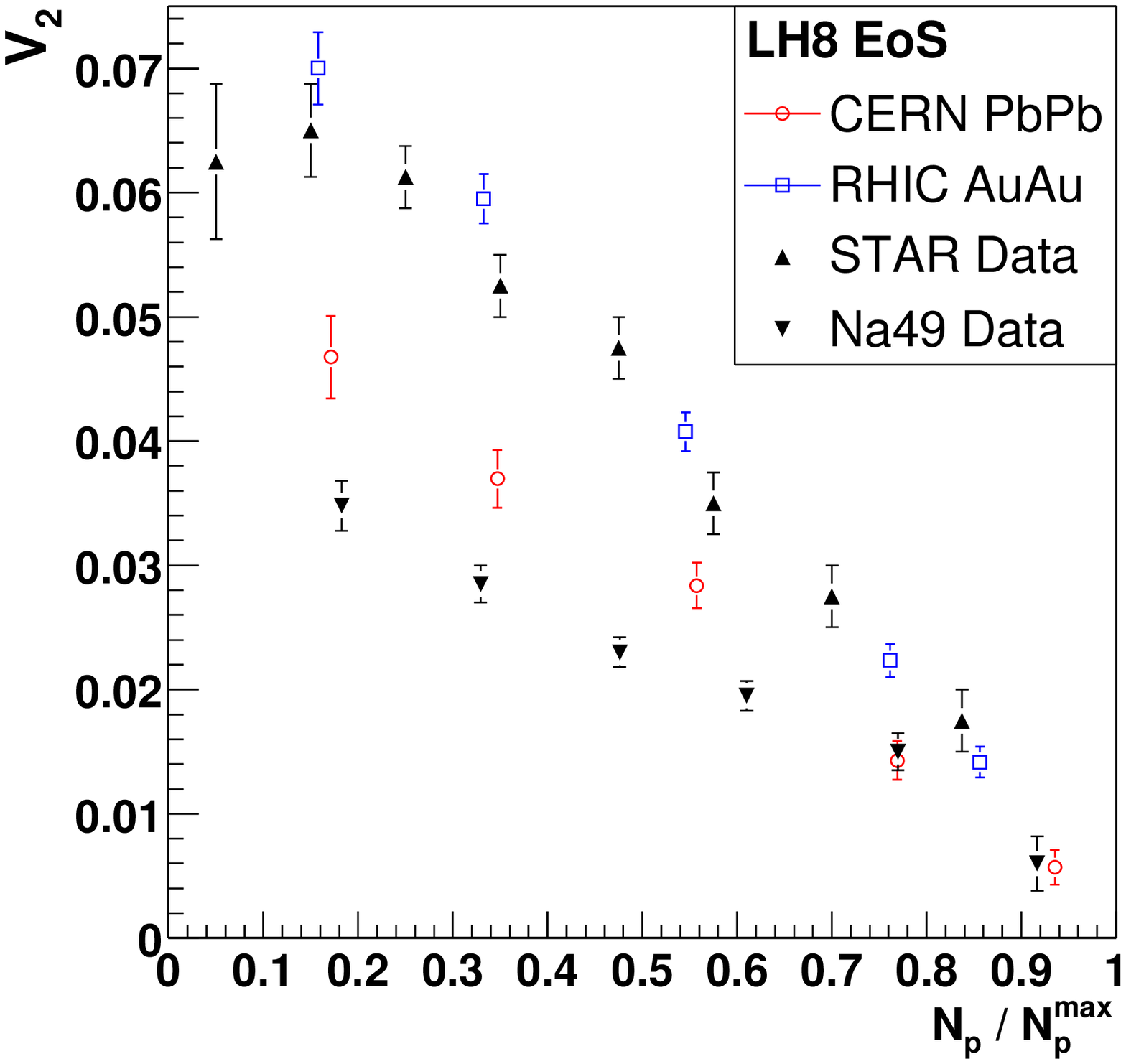}
\end{center}
   \caption[]{
      \label{psPionV2dNdy}
(a) The  elliptic flow  
parameter $V_2$ versus multiplicity at b=6 fm, for different EoS. (b)
  $V_2$ is now plotted
 versus impact parameter b, described experimentally by the
number of participant nucleons, for RHIC STAR and SPS NA49 experiments. 
 Both are compared to our results for EoS LH8.
   }
\end{figure}

Fig.2b shows   data as a function of impact parameter. One can see
that the agreement becomes much better at RHIC. Furthermore, one may
notice that deviation from linear dependence we predict
becomes visible at SPS for more peripheral
collisions with $N_p/N_p^{max}<0.6$ or so, 
while at RHIC only the most peripheral point, with
$N_p/N_p^{max}=0.05$
show such deviation. This clearly shows that hydrodynamical regime in
general
works much better at RHIC. On the other hand, a models of single
re-scattering (e.g. 
\cite{Heiselberg}) which approximately describes peripheral SPS data is
completely inconsistent with this linear rise, seen both at SPS and 
RHIC.

\section{Matter composition}
\subsection{Strangeness}
The discussion of the role of s,c quarks in excited hadronic matter
has a peculiar history. In the first paper addressing the subject
\cite{Shu_QGP} it was argued that QGP with $T\sim 500 \, MeV$ should
quickly
equilibrate even charm, because of high rate of gluon-induced
processes,
and suggested to use it as a QGP signature. As for strangeness, 
the $m_s$ was considered to be too small to be useful for that
purposes.  However 
 later, Rafelski et al \cite{KMR}
had nevertheless suggested that the 
strangeness enhancement  for exactly that
role.
The experiments at AGS/SPS indeed found lager strangeness per
participant
baryon (or per pion) in AA collisions than in pp, which is more
pronounced
for hyperons and especially for $\Omega^-$. However, the following two
features has been also found. First, strangeness enhancement
started at very low energies, at AGS. Second, its magnitude basically
described by the statistical model (see e.g.\cite{stat}),
 by an equilibrium
at $T\approx 170 \, MeV$ and $\mu_b$ depending on the
collision energy. From this point
of view people rather called a phenomenon a ``disappearance of
strangeness suppression'', which has been
present in pp.  

But even this interpretation has been recently challenged in \cite{Redlich}. Already
25 years ago the issue of correct account for strangeness in small
statistical systems has been addressed in \cite{Shu_oldtherm}. The main point is,
the usual statistical expressions say for $K/\pi$ ratio, being the
ratio of integrated Bose distribution, is only valid if the 
average particle number $<N_K> \gg 1$. If it is not so, exact
strangeness
conservation should be enforced while considering all possible states
of the system, say containing $KK$ pairs, etc.  As a result, $<N_K>$
depends on volume (or $<N_\pi>$) quadratically, till    $<N_K> \sim 1$.
In  $\bar p p, pp$ collisions at few GeV range the data were in
perfect agreement with this prediction already in 1975 \cite{Shu_oldtherm}.
We are however still lacking a demonstration of where that happens
for say $\Xi,\Omega$: even most peripheral data at SPS do not show
 a predicted transition to small-volume regime, their ratio to
multiplicity
remains flat. Probably lighter ions are needed to see it.

In summary: whatever strange it may appear to us, composition
of hadrons, including strange ones, seem to be thermal $both$ in AA (at
AGS/SPS energies) and pp collisions. In the latter case it is believed
to
come from string fragmentation: so AA  can either be also
string-based, or come from QGP, which (by coincidence?) have a
$T_c$ value which mimic string decays.

\subsection{ Solving the anti baryon puzzle}
  Anti-baryon ratios, like many others,  
can be rather accurately characterized by the so-called {\em chemical} 
freeze-out stage with a 
common temperature $T_{ch}$ $\mu_{B,ch}$  (values depend
on collision energy). However, the kinetics of other particles and
anti-baryons
cannot be the same. The number of
pions, kaons, etc. are not subject to significant changes when 
the system evolves from $T_{ch}$ to $T_{th}$: thus
{\em 'over-saturation'} appears, the effective  
pion  fugacity
$$z_\pi=exp(\mu_\pi/T_{th})\approx 1.6-1.9$$
The situation for antibaryons is different because
 the pertinent annihilation cross section is 
 {\em not small}, $\sigma_{p\bar p\to n\pi} \simeq 50$~mb.
The time in which a give antinucleon is eaten is only 
$$\tau_{ch}=\frac{1}{\sigma_{ann} \ \rho_B \ v_{th}}\simeq 3~{\rm
fm/c} $$
and so naively one might expect most of the antiprotons to be annihilated, and 
various transport calculations ( ARC, UrQMD)
have indeed  been unable to account for the measured number, falling short
by  significant factors. Speculations have been put forward:
either a reduction of the annihilation rate
or an enhanced production.

 Inverse reactions, ignored in event generators, are however
{\em not small} \cite{RS_anti} 
$${\cal R}_{th} = |{\cal M}_n|^2\exp[-(E_p+E_{\bar p})/T] \ \left[z_p \ z_{\bar p} \ 
 - 
z_\pi^{ n} \right] $$
The condition   that this reaction goes on
implies  \cite{RS_anti}
$$
z_{\bar p} = (z_p)^{-1} {\langle z_\pi^n\rangle}
$$
which leads to predictions consistent with
 the experimental value of $N_{\bar p}/N_p=(5.5\pm 1)\%$
 (na44). Note that it is very multi-pion reaction, with n=6-7.

After the paper by Rapp and myself,
 C.Greiner and S.Leupold \cite{GL} 
generalized it further, to $\bar Y+N \leftrightarrow n\pi+K$ where
$Y=\Sigma\Xi,\Omega$. Similar multi-meson annihilation
can also explain the long-standing puzzle about
multiple production {\em anti-hyperons}.

\subsection{Charm and  $J/\psi$ suppression}
 Charm production at SPS is due to hard gg collisions,
which are obviously able to produce somewhat larger amount
 of it, compared to its equilibrium value at $T=170 \, MeV$.
Direct measurements are still in the future, but 
we should hardly expect any surprises here. (The NA50 medium mass dilepton
excess fits nicely to thermal dilepton rate \cite{RS_dil}, so I do not 
think it is a charm enhancement.)

Do $J/\psi,\psi'$ we observe in AA collisions know about hadronic matter
which surrounds it, or their cross section is too small and too
absorptive (leading to $\bar D D$ pairs) that all of them we see just jump
intact directly from the primary production point? The latter remains
the prevailing view in the field, although it has been challenged
lately.

Although  $\psi'/J/\psi$ ratio starts decreasing at rather peripheral
collisions/small A, as larger radius of  $\psi'$ suggests larger cross
section,
it then stabilizes at  $\psi'/J/\psi\approx 0.05$. Furthermore, 
as noticed in \cite{SSZ}, this number is consistent with {\em the same} 
  $T\approx 170 \, MeV$ which fits all other ratios. This leads to an
idea that the observed $\psi'$ is  excited from $J/\psi$.

Much more radical idea has been put forward in \cite{PBM_charm}: the
number
of  $J/\psi$ itself is in agreement with the statistical model, provided
the total charm fugacity is fixed to the parton model production. 
If it is not a coincidence, it  then implies that 
$J/\psi$ may in fact be created from the same heat bath as all other
hadrons. If so, the individual string scenario is of course out of the
window, and the QGP scenario (with some extra charm added to it) is
in.

Much more 
charm is expected at RHIC: it makes recombination of floating charm
pairs into charmonium states even more likely. The issue has been
studied in recent paper \cite{RT}: the conclusion is that we should
in fact expect a {\em charmonium enhancement} at RHIC.

\section{Event-by-event Fluctuations}
\subsection{Current SPS data versus theory}
Unlike pp collisions, the AA ones demonstrate small fluctuations of Gaussian
shape, as measured e.g. by NA49. It is basically a consequence of
a basic theorem of probability theory, roughly stating that any
distribution with a maximum in high power becomes Gaussian.
Still, it is quite striking how far this Gaussian goes: no 
 deviations from it is seen for few decades, as far as data go.
No trace of any large fluctuations, DCC or other bubbles, and all
events with the same impact parameter
are basically alike.   

The question then is whether the $widths$ of those Gaussians is understood.
 The short answer is yes. Somewhat longer answer is that one should
separate \cite{SRS} fluctuations of two types, of intensive or extensive
 variables.
The latter is e.g. total charge multiplicity: it obtains about equal
contributions from the initial  (due to fluctuations in spectators) and final 
stage (resonances). The former (e.g. particle ratios like
 $\pi^+/\pi^-$) are well described by resonances at freezeout.
 
To give some example, consider fluctuations in charged multiplicity
measured by NA49 at 5\% centrality:$ {<\Delta N_{ch}^2> \over
<N_{ch}>}\approx 2.0$,  while Poisson statistics (independent production) gives 1 in the rhs.

Statistical fluctuations in a resonance gas increase the rhs. Example:
if all pions would originate from $\rho^0\rightarrow
\pi^+\pi^-$, and rho would be random, 
then the rhs would be 2. Account for equilibrium composition lead
however
to the Gaussian width of only about
1.5 \cite{SRS}.  Correlations coming from fluctuation in the number of
participants further increase it \cite{DS}: see the comparison
in the figure \ref{fig_fluct}: 
\begin{figure}[t]
\epsfxsize=3.5in
\centerline{\epsffile{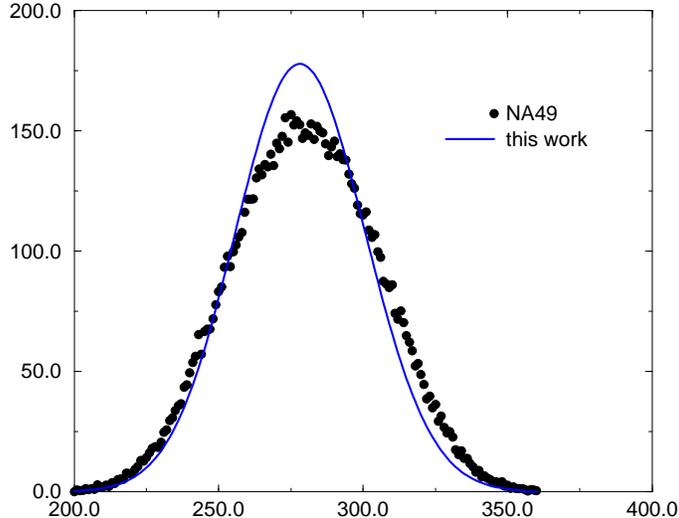}}
\vskip -0.1in
\caption[]{
 \label{fig_fluct} The calculated multiplicity distribution from  our
 model (solid line) is compared to the observed one
(points).  }
\end{figure}
Good examples of intensive variable  fluctuations are those  in the
mean $p_t$ (in an event) or in electric charge Q (same as
of the $\pi^+/\pi^-$ ratio). Resonances again: e.g. $\rho^0\rightarrow
\pi^+\pi^-$ does not lead to Q but contribute to multiplicity
$ Q^2/<N_ch>\approx .7-.8$, not 1 as for Poisson. 
It is what is actually observed for central collisions by NA49.
\subsection{
Can ``primordial'' fluctuations be seen?}
In order to do so, we need a telescope looking back into the past,
through the clouds of hadronic matter. How to make it has been
recently suggested in refs \cite{fluct_new}. 
 The idea is to use {\em long-wavelength fluctuations of  conserved
quantities}, which have slow relaxation. Example: primordial fluctuations in
microwave background give us ``frozen plasma oscillations'' at Big
Bang.
 Can we find similar signal in the ``Little Bang''?
Yes: if relax.time is longer than lifetime of
hadronic stage the fluctuations we would see are ``primordial
ones'', from the QGP which are factor $\sim 3$ smaller.

Quantitative studies have been recently done by Stephanov and myself
\cite{ourfluct_new}.
 Fluctuations are governed by
Langevin eqn $$  {\partial_\tau f} = \gamma(\tau)\, \partial^2_y f + \xi(y,\tau)$$
and we calculated $\gamma(\tau)$ for pions in collisions. Basically we 
found that at RHIC/SPS pion diffuse during hadronic stage by about 2
units of rapidity. NA49 data we had are in acceptance $\Delta y=2$ and
show perfect agreement with equilibrium resonance gas.   STAR detector
may have  $\Delta y=4$ and additional deviation from 1 by 
about 20\% - quite observable.

 Baryon number fluctuations idea cannot work because we  do
not see neutrons.
\section{Summary}
\begin{itemize}
\item{--} 
If hydro description plus lattice-like EoS makes sense, we found that
the QCD phase transition plays different role at AGS/SPS and
RHIC: it makes the EoS effectively soft in the former case but
stiff at high energy, providing early ``QGP push''.
\item{--} Collective flow, especially its elliptic component,
is very robust measure of the pressure in the system. STAR data at
RHIC, which demonstrate increase of elliptic flow by factor 2,
contradicts
to string-based models and also mini-jet models without rescattering.
Hydro-based model, especially H2H model with hadronic cascades,
can describe these data very accurately. Still two solutions seem
to be possible, one with strong QGP push (predicted by lattice-based
EoS) and another with longer-time burning.
\item{--} Strangeness seem to be well equilibrated in any collisions,
and thus is not a QGP signature. Whether    $J/\psi$
can obtain significant contribution from equilibrated charm at SPS is 
strongly debated: it should however be the case at RHIC.
\item{--} Event-by-event fluctuations are mostly due to final state
interaction between secondaries, and available data are in agreement
with equilibrium resonance gas calculations.

\item{--} Potential observations of ``primordial charge fluctuations''
are limited by their relaxation time, mostly
due to pion diffusion in rapidity. The outlook depends on experimental
acceptance: available calculations suggest that wide STAR acceptance
can be sufficient to see 10-15 percent modification of charge fluctuations.

\end{itemize}

\end{document}